\documentclass[12pt,a4paper]{article}
\usepackage[english]{babel}
\usepackage{graphicx}

\newcommand{\alt}{\mathbin{\lower 3pt\hbox
   {$\rlap{\raise 5pt\hbox{$\char'074$}}\mathchar"7218$}}}
\newcommand{\agt}{\mathbin{\lower 3pt\hbox
   {$\rlap{\raise 5pt\hbox{$\char'076$}}\mathchar"7218$}}}

\begin{document}
\setcounter{footnote}{0}
\setcounter{equation}{0}
\setcounter{figure}{0}
\setcounter{table}{0}
\vspace*{5mm}



\begin{center}
{\large\bf Renormalization Group Functions of the $\phi^4$ \\
Theory from High-Temperature Expansions }

\vspace{4mm}
I. M. Suslov\\
Kapitza Institute for  Physical Problems,\\
 Moscow, 119334 Russia\\
 E-mail: suslov@kapitza.ras.ru
\vspace{1mm}
\end{center}

\begin{center}
\begin{minipage}{135mm}
{\bf Abstract } \\
It has  been previously shown that calculation of  renormalization group
 (RG) functions of scalar $\phi^4$  theory is reduced to
 thermodynamic properties of the Ising model. Using
 high temperature expansions for the latter, RG functions of the
 four-dimensional theory can be calculated for arbitrary coupling
 constant $g$,  with an accuracy   of $10^{-4}$ for the
 $\beta$-function and with an accuracy of $10^{-3}$ --
 $10^{-2}$ for anomalous dimensions. The expansions of the
 RG functions up to the 13th order in
 $g^{-1/2}$ have been obtained.

\end{minipage}
\end{center}

\vspace*{1.5mm}
\vspace*{1.5mm}

\vspace{6mm}
\begin{center}
{\small\bf 1. INTRODUCTION}
\end{center}

As was recently shown in [1, 2], the Gell-Mann -- Low function
$\beta(g)$ and anomalous dimensions of the $\phi^4$ theory can
be expressed in terms of the functional integrals, providing
the representation
$$
g=F(g_0,m_0,\Lambda)\,,\qquad
\beta(g)=F_1(g_0,m_0,\Lambda)\,,
\eqno(1)
$$
where $g_0$ and $m_0$ are the bare charge and mass,
respectively; $\Lambda$ is the momentum cutoff parameter, and
$g$ is the renormalized charge. Large $g$ values are reached
only near a zero of one of the functional integrals, where the
right-hand sides of Eqs.1 are significantly simplified and the
parametric representation is resolved in the explicit form. As a
result, asymptotic expressions for the $\beta$ function and
anomalous dimensions are obtained. A similar approach can also
be implemented in QED [3].

Parametric representation (1) has the
following general property. If $g_0$ is expressed in terms of $g$
using the first of Eqs.1 and the resulting expression is
substituted into the second equation, the dependence on $m_0$ and
$\Lambda$  disappears according to the general theorems [4], so
that the $\beta$  function depends only on $g$.  However, this
property is not automatically satisfied in applied calculations.
The reason is that the general theorems imply the continual limit
$\Lambda\to\infty$, which physically means the condition
$$
m\ll \Lambda \qquad
\mbox{ or}  \qquad
\xi \gg a\,,
\eqno(2)
$$
where $m$ is the renormalized mass, $\xi$ is the correlation
radius, and $a=\Lambda^{-1}$ is the constant of a lattice at
which the functional integral is defined.  Under condition (2)
in the region of large $g_0$ values, the functional integrals
of the $\phi^4$ theory are reduced to Ising sums; as a
result, Eqs.1 have the form
$$
g= F(\kappa)\,,\qquad
\beta(g)= F_1(\kappa)\,
\eqno(3)
$$
where $\kappa$ has the meaning of inverse temperature in the
Ising model, and it is obvious that the $\beta$ function
depends only on g. Condition (2) formally corresponds to the
inequality
$-g_0^{-1} m_0^2/\Lambda^2 \agt 1$,
but the reducing to the Ising model is really possible
under the weaker condition
$$
 g_0\gg 1\,,\qquad -g_0^{-1/2} m_0^2/\Lambda^2 \gg 1\,,
\qquad -g_0^{-1} m_0^2/\Lambda^2 =\kappa
 \mbox{ --- arbitrary}\,.
\eqno(4)
$$
For this reason, parametric representation (3) remains valid
in the region of small $\kappa$, where  values of $g$
are large and gradient expansions are applicable. At first
glance, the condition $g_0\gg 1$ corresponds to the strong
coupling regime and parametric representation (3) is limited
by only this condition. However, there is another view on this
situation. Let us strengthen conditions (4) by passing to the
limit
$$
g_0\to\infty\,,\qquad -g_0^{-1/2} m_0^2/\Lambda^2 \to \infty\,,
\qquad -g_0^{-1} m_0^2/\Lambda^2 =\kappa = const
\eqno(5)
$$
In this case, the transition from Eqs.1 to Eqs.3
is valid without any approximations and conserves
strict equivalence with the initial $\phi^4$ theory under a
certain choice of its bare parameters; the last property
ensures the conservation of the form of the Lagrangian
under renormalizations. The passage to the limit $g_0\to\infty$
does not mean the same passage for the renormalized charge $g$;
in fact, according to gradient expansions, $g$ varies from
infinity to about unity when $\kappa$ varies from zero to about
unity.  Since parametric representation (3) is exact and
specifies the $\beta$  function in the interval $1\alt  g
<\infty$, it can be analytically continued and treated as a
definition of $\beta(g)$ at arbitrary $g$ values. However,
there is a question: Does this definition provide correct
results in the weak-coupling region?

An answer to this question can be obtained using high-temperature
series [5]. Such series are traditionally constructed for
quantities
$\raisebox{2pt}{$\chi$}_2$, $\raisebox{2pt}{$\mu$}_2$,
$\raisebox{2pt}{$\chi$}_4$
(see Section 2), which completely specify the right-hand
sides of Eqs.3. High-temperature expansions are formally
applicable for small $\kappa$, but their comparatively
large length (up to 30 terms in some cases) allows a successful
analysis of the vicinity of the phase transition point
$\kappa_c$ and leads to the results consistent with other
methods. Consequently, good approximations for the indicated
quantities can be obtained
throughout the interval $0\le \kappa\le \kappa_c$. The
substitution of such results into the right-hand sides of
Eqs.3 makes it possible to determine the renormalization group
functions in the interval $ g^*\le  g <\infty$, where $g^*$ is the
fixed point of the renormalization group. In the four-dimensional
case, $g^* =0$ and the mentioned procedure completely determines
the renormalization group functions. In many works [6 --- 16],
the high-temperature series were used to test logarithmic
corrections to scaling [17].  Already those works provide the
positive answer to the above question:  parametric representation
(3) gives correct results in the weak-coupling region.
Therefore, we can concentrate our efforts on constructing the
renormalization group functions of the four-dimensional
$\phi^4$ theory for arbitrary $g$ values. This can be done with
an accuracy of $10^{-4}$ for the $\beta$ function and with a
slightly lower accuracy for anomalous dimensions.

The determination of calculated renormalization group functions
implies the use of a lattice regularization different from the
usual Pauli--Villars regularization scheme, isotropic cutoff in
the momentum space, dimensional regularization, etc. However, the
$\beta$ function in the used scheme is determined in terms of
the observed charge and mass [1, 2] and should be independent of
the cut-off procedure. Such a dependence is possible for
anomalous dimensions, because they are determined in terms of the
unobservable $Z$ factors. In any case, the distinction of this
way of regularization from the usual  procedures is no more
than difference between the latter procedures

\vspace{6mm}
\begin{center}
{\small\bf 2. INITIAL RELATIONS}
\end{center}

Let us consider the $n$ component $\phi^4$
theory with the action
$$
S\{\varphi\} =\int \,d^dx \left\{
{\textstyle\frac{1}{2}} \sum_{\alpha=1}^n
(\nabla \varphi_\alpha)^2
+ {\textstyle\frac{1}{2}} m_0^2
\sum_{\alpha=1}^n \varphi_\alpha^{\,2} +
{\textstyle\frac{1}{8}} u_0
\left(\sum_{\alpha=1}^n\varphi_\alpha^{\,2}\right)^2
\right\}\,,
$$
$$
u_0=g_0\Lambda^{\epsilon}\,, \qquad \epsilon=4-d\,,
\eqno(6)
$$
where $g_0$ and $m_0$ are the bare charge and
 mass, respectively; $d$ is the dimensionality of space; and
 $\Lambda$ is the momentum cutoff parameter. The most general
 functional integral of this theory contains $M$ multipliers
 of  the field $\phi$ in the pre-exponential factor,
$$
Z^{(M)}_{\alpha_1\ldots \alpha_M}(x_1,\ldots, x_M)=
\int D\varphi\,
\varphi_{\alpha_1} (x_1) \varphi_{\alpha_2} (x_2) \ldots
\varphi_{\alpha_M} (x_M) \exp\left(-S\{\varphi \} \right) \,,
\eqno(7)
$$
and will be denoted as $K_M\{p_i\}$  after
 the transition to the momentum representation and the separation
 of $\delta$ factors,
$$
Z^{(M)}_{\alpha_1\ldots \alpha_M}(p_1,\ldots, p_M)=
K_M\{p_i\}\, {\cal N}\,\delta_{p_1+\ldots+p_M}
I_{\alpha_1\ldots \alpha_M}\,,
\eqno(8)
$$
where $I_{\alpha_1\ldots \alpha_M}$ is the sum of terms
$\delta_{\alpha_1 \alpha_2} \delta_{\alpha_3 \alpha_4}\ldots$
 with all possible pairings, and ${\cal N}$ is the number of
 sites of the lattice on which the functional integral is defined.
The integrals $K_M\{p_i\}$  are usually estimated at zero momenta
and  only one integral $K_2\{p\}$  is required for small $p$
values,
$$
 K_2(p)=K_2-\tilde K_2
p^2+\ldots
\eqno(9)
$$
Below, the case with $d=4$ and $n=1$ is considered,
but the general formulas are written for arbitrary $d$ and
 $n$ values.

 The below consideration concerns the renormalization
 group functions $\beta(g)$, $\eta(g)$, and $\eta_2(g)$
 entering into the Callan--Symanzik equation  \cite{4}
$$
\left[ \frac{\partial}{\partial\ln m} +
\beta(g) \,\frac{\partial}{\partial g}
+\left(L-N/2 \right) \eta(g) -L\eta_2(g)
\right] \Gamma^{(L,N)} = 0\,,
\eqno(10)
$$
for the  vertex  $\Gamma^{(L,N)}$  with $N$ external lines of the
field $\phi$ and $L$ external interaction lines.  The expression
of these functions in terms of the functional integrals leads to
the parametric representation  \cite{2}
$$
g=-\left(\frac{K_2}{\tilde K_2} \right)^{d/2}
\frac{K_4 K_0}{K_2^2} \,,
\eqno(11)
$$
$$
\beta(g)=-\left(\frac{K_2}{\tilde K_2} \right)^{d/2}
\frac{K_4 K_0}{K_2^2} \,
\left\{ d +2 \,\frac{(\ln K_4 K_0/K_2^2)'  }
{ (\ln K_2/\tilde K_2)' }
\right\}   \,,
\eqno(12)
$$
$$
{    }
$$
$$
\eta(g)= 2\, \frac{(\ln K_2/K_0)'+(\ln K_2/\tilde K_2)'  }
{ (\ln K_2/\tilde K_2)' }\,,
\eqno(13)
$$
$$
{    }
$$
$$
\eta_2(g)=-2\, \frac{ (\ln K_0/ K_2)'' +
\left[(\ln K_0/ K_2)'\right]^2 }
{(\ln K_2/\tilde K_2)'\, (\ln K_0/ K_2)' }\,,
\eqno(14)
$$
where primes stand for derivatives with respect to $m_0^2$.
Under condition (4), the functional integral of the scalar theory
can be written in the form
$$
Z_M\{{\bf x}_i\}=(2\kappa)^{\frac{{\cal N}+M}{2}}
    \int \left(\prod_{\bf x}\,d\varphi_{\bf x} \right)
      \varphi_{{\bf x}_1} \ldots \varphi_{{\bf x}_M}
\exp\left\{- \kappa\,\sum_{\bf x,x'}
J_{\bf x-x'} \varphi_{\bf x}\varphi_{\bf x'} \right\} \prod_{\bf
 x} \delta(\varphi^2_{\bf x}-1)
\eqno(15)
$$
and is transformed to an Ising sum over the values
$\varphi_{\bf x}=\pm 1$. The quantities studied in
high-temperature expansions are introduced as
$$
\raisebox{2pt}{$\chi$}_2=\sum_{\bf x} \langle \varphi_{{\bf
x}}\varphi_{{\bf 0}} \rangle^c\,,\qquad
\raisebox{2pt}{$\mu$}_2=\sum_{\bf x} {\bf x}^2\langle
\varphi_{{\bf x}} \varphi_{{\bf 0}} \rangle^c\,,\qquad
\raisebox{2pt}{$\chi$}_4=\sum_{\bf x,y,z} \langle
\varphi_{{\bf x}} \varphi_{{\bf y}}\varphi_{{\bf z}}\varphi_{{\bf
0}} \rangle^c\,,\qquad
\eqno(16)
$$
(where superscript $c$ marks the connected diagrams)
and coincides up to factors with the ratios $K_2/K_0$,
$\tilde K_2/K_0$, and $K_4/K_0$ of the functional integrals
introduced above; more precisely,
$$
\frac{K_2}{\tilde K_2}\,=\,2d\, \frac{\raisebox{2pt}{$\chi$}_2}
{\raisebox{1pt}{$\mu$}_2}\, \equiv
\frac{1}{\kappa} f_0(\kappa) \,,
$$
$$
\frac{K_2}{K_0}\,=\,2\kappa\,\raisebox{2pt}{$\chi$}_2
\equiv \kappa f_2(\kappa) \,,
\eqno(17)
$$
$$
\frac{K_4 K_0}{K_2^2}\,=\,\frac{1}{3}\,
\frac{\raisebox{2pt}{$\chi$}_4}{\raisebox{2pt}{$\chi$}_2^2}
\,\equiv - f_4(\kappa) \,,
$$
where the introduced functions $f_i(\kappa)$ will be used below.
It was taken into account that there is no zeroth term in the
expansion of ${\raisebox{1pt}{$\mu$}_2}$ in $\kappa$
(see Eq.20 below), so that all functions $f_0(\kappa)$,
$f_2(\kappa)$, and $f_4(\kappa)$ are regular and their
expansions begin with the zeroth term. The substitution of
Eqs.\,17 into Eqs.\,11\,--\,14
gives
$$
g=\left(\frac{f_0(\kappa)}{\kappa} \right)^{d/2}
f_4(\kappa) \,,
$$
$$
\frac{\beta(g)}{g}= d - 2\kappa \,\frac{[\ln f_4(\kappa)]'  }
{ 1-\kappa \,[\ln f_0(\kappa)]' }
   \,,
$$
$$
{    }
$$
$$
\eta(g)= - 2\kappa\, \frac{[\ln f_0(\kappa) f_2(\kappa)]'  }
{ 1-\kappa\, [\ln f_0(\kappa)]' }\,,
\eqno(18)
$$
$$
{    }
$$
$$
\eta_2(g)=-2\, \frac{\left(1+\kappa\, [\ln f_2(\kappa)]'\right)^2
+1 - \kappa^2\, [\ln f_2(\kappa)]''  }
{\left(1-\kappa\, [\ln f_0(\kappa)]'\right)\,
\left(1+\kappa \,[\ln f_2(\kappa)]'\right) }
$$

It is easy to obtain the strong coupling behavior for
renormalization group functions taking limit $\kappa\to 0$
\cite{2}:
$$
\beta(g)=d g\,,\qquad \eta(g)=0\,,\qquad \eta_2(g)=-4\,\qquad
(g\to \infty)\,.
\eqno(19)
$$
For a simple hypercubic lattice with the
interaction between the nearest neighbors, the first terms of
the expansion of functions (16) for $d = 4$ and $n = 1$ have
the form  \cite{17}
$$
\raisebox{2pt}{$\chi$}_2= 1 +16 \kappa + 224 \kappa^2 + \ldots
$$
$$
\raisebox{1pt}{$\mu$}_2= 16 \kappa + 512 \kappa^2 +
33920/3 \kappa^3 +\ldots
\eqno(20)
$$
$$
\raisebox{2pt}{$\chi$}_4= -2 - 128 \kappa - 4672 \kappa^2 -
\ldots
$$
The substitution into Eqs.\,18 makes it possible
to obtain the expansion of the renormalization group
functions in  $g^{-2/d}$ and, in particular, a more accurate
asymptotic expression for  $\eta(g)$
$$
\eta(g)= \frac{16}{9}\,\frac{1}{ g}\,,\qquad
g\to\infty \,.
\eqno(21)
$$
The universality of this asymptotics  has
not been tested and, strictly speaking, it refers
to the indicated model. Below, 14 terms of expansion (20)
presented for $n = 1$ in tables 5, 8, and 11 of the paper
[18] are used.

\vspace{3mm}
\begin{center}
{\small\bf 3. VICINITY OF THE PHASE TRANSITION}
\end{center}

\begin{center}
{\small\bf 3.1. General Strategy }
\end{center}

The foundation of the application of high-temperature
expansions for investigating the critical behavior is
as follows. Let a certain quantity $F(\kappa)$  has
a power-law behavior near the transition point
$\kappa_c=1/T_c$
$$
F\propto (T-T_c)^{-\lambda} \propto
(\kappa_c-\kappa)^{-\lambda} \,.
\eqno(22)
$$
In this case,  the  convergence  radius  of  the expansion in
$\kappa$ is limited by the quantity $\kappa_c$. In actual
cases, $\kappa_c$ is the nearest
singularity to the coordinate origin; this circumstance
facilitates its analysis. It is easily seen that the nearest
singularity for the logarithmic derivative
$$
(\ln F)' =\frac{F'}{F} \sim \frac{{-\lambda}}{\kappa-\kappa_c}
\eqno(23)
$$
is a simple pole with a residue ${-\lambda}$ and can be
investigated using the Pade-approximation. The Pade-approximant
$[M/N]$ is defined as the ratio of the polynomials of the degrees
$M$ and $N$,
$$
(\ln F)' = \,\frac{P_M(\kappa)}{Q_N(\kappa)}\,= \,
\frac{p_0+p_1 \kappa+\ldots +p_M \kappa^M}
{1+q_1 \kappa+\ldots +q_N \kappa^N} \,\,,
\eqno(24)
$$
whose coefficients are chosen such that the first
$M +N+ 1$ coefficients of the expansion of $(\ln F)'$ in $\kappa$
are reproduced.  It is known that Pade-approximants
successfully predict the nearest singularities of the
approximated function if these singularities are simple poles
[5, 19]. Diagonal ($M = N$) or quasidiagonal ($M\approx N$)
approximants are usually used for which convergence to the
corresponding function is proved under the most general
assumptions. The use of this strategy in the four-dimensional
case is complicated by the existence of logarithmic corrections
to scaling \cite{19,4}:
$$
\raisebox{2pt}{$\chi$}_2 \sim \tau^{-1} |\ln \tau|^p \,,
\qquad\qquad\qquad\qquad\qquad\qquad\qquad\qquad
$$
$$
\xi^2 \sim \frac{\raisebox{1pt}{$\mu$}_2}
{\raisebox{2pt}{$\chi$}_2}
 \sim \tau^{-1} |\ln \tau|^p\,,
\qquad\qquad\qquad p=-\frac{\zeta_1}{\beta_2}=\frac{n+2}{n+8}
\eqno(25)
$$
$$
\raisebox{2pt}{$\chi$}_4 \sim \tau^{-4} |\ln \tau|^{4p-1}\,,
\qquad\qquad\qquad\qquad\qquad\qquad\qquad\qquad
$$
where $\tau\sim (\kappa_c-\kappa)$ is the distance to the
transition and the exponent $p$ is determined by the first terms
of the expansion of the renormalization group
functions,
$$
\beta(g)=\beta_2 g^2 +\beta_3 g^3 +\ldots\,,
$$
$$
\eta(g)=\delta_2 g^2 +\delta_3 g^3 +\ldots \,,
\eqno(26)
$$
$$
\eta_2(g)=\zeta_1 g +\zeta_2 g^2 +\ldots \,,
$$
where
$$
\beta_2=S_4 \,\frac{n\!+\!8}{2}\,,\qquad
\beta_3=-S_4^2 \,\frac{9n\!+\!42}{4}\,,\qquad
\delta_2=S_4^2 \,\frac{n\!+\!2}{8}\,,\qquad
\zeta_1=-S_4 \,\frac{n\!+\!2}{2}\,\qquad
\eqno(27)
$$
and $S_4=1/8\pi^2$. According to (25) we have for functions
 $f_i$
$$
f_0 \sim \tau |\ln \tau|^{-p} \,,\qquad
f_2    \sim \tau^{-1} |\ln \tau|^p \,,\qquad
f_4 \sim \tau^{-2} |\ln \tau|^{2p-1} \,.\qquad
\eqno(28)
$$
The behavior of the charge $g$ is given by the expression
$$
g=\frac{c_0}{|\ln \tau|} \,, \qquad c_0=2/\beta_2 \qquad
(\tau\to 0)  \,,
\eqno(29)
$$
where the coefficient of the logarithmic factor is universal.
 When Eqs. (28) and (29) are valid, parametric representation
 (18) automatically ensures the results
 $\beta(g)=\beta_2 g^2$, $\eta(g)=0\cdot g$,
and $\eta_2(g)=\zeta_1 g$, the correct behavior of the
 renormalization group functions at small $g$ values.

The objective test of Eqs.25) for lattice models were performed
in many works [6--15]. In particular, it was convincingly shown
in [6, 7] that high-temperature series for the Ising model allow
reliable prediction of the exponent $p$. Expression (29) was
confirmed with a satisfactorily accuracy in [7, 9]. Already
these results provide the positive answer to the question
formulated in the Introduction:  parametric representation
(18) gives correct results for the renormalization group
functions in the weak-coupling region.

\begin{center}
{\small\bf 3.2.  Zeroth Approximation }
\end{center}

The Pade-analysis of Eqs.\,28 is performed by
the successive approximation method. In the zeroth approximation,
the logarithmic factors are ignored  and the functions $f_i$ are
processed under the assumption of their power-law dependence
on $\kappa$. The results of such an analysis presented in Table 1
show a significant difference of the obtained exponents from the
exact values (see Eqs.\,28) and provide a rough estimate of
the critical point
$$
\kappa_c= 0.07476 \div 0.07490
$$
A more accurate estimate of $\kappa_c$
can be obtained taking into account that the ratio
$ \raisebox{2pt}{$\chi$}_4/\raisebox{2pt}{$\chi$}_2\sim
f_4 f_2 $ in the scalar case (when $p = 1/3$) behaves as $\tau^{-3}$
and  contains no logarithms [6]. As is seen in Table 2, the
Pade-analysis of this quantity provides the exponent
really close to the exact value and the corresponding
estimate of $\kappa_c$
$$
\kappa_c= 0.07481 \div 0.07487
\eqno(30)
$$
is  almost final and will be only slightly refined below.
The central  value of interval (30) almost coincides
with the result $\kappa_c= 0.074834(15)$ obtained in \cite{6}
 with a more sophisticated  processing.

\begin{center}
{\small\bf 3.3. First Approximation}
\end{center}

In this approximation, the  following representation is
 used:
$$
f_0 =\tilde f_0 |\ln \tau|^{-p} \,,\qquad
f_2 =\tilde f_2   |\ln \tau|^p \,,\qquad
f_4 =\tilde f_4  |\ln \tau|^{2p-1} \,\qquad
\eqno(31)
$$
and the Pade-analysis is applied to the functions $\tilde f_i$.
Since the relation $\tau=A(\kappa_c-\kappa)$ includes
the nonuniversal factor $A$, it can be accepted that
$$
|\ln \tau| = A_0 - \ln (1-\kappa/\bar\kappa_c)\,,
\eqno(32)
$$
\begin{center}
\hspace{10mm} {\bf  Table 1.}
 Position of the pole corresponding to the critical point
$\kappa_c$ and residue at it (in parentheses)
for the Pade approximant $[N/N]$  of functions
$[\ln f_i(\kappa)]'$.\,\footnote{\,The
asterisk in  Tables 1--4 marks defective
approximants.  A "defect"  in the Pade-analysis is the appearance
 of a pair of a pole and a root close to each other; as a result,
 the corresponding Pade-approximant is reduced to a lower order
 approximant. The defectiveness of the approximant can lead to
 loose of the accuracy and is a reason for its
 discrimination.}
 \vspace{2mm}

\begin{tabular}{||c|c|c|c||}
\hline
 & &  &    \\
$N$     &  $[\ln f_0(\kappa)]'$  & $[\ln f_2(\kappa)]'$
&  $[\ln f_4(\kappa)]'$\\
  & & & \\
  \hline
  & & & \\
 2 & 0.07519 (1.130) & 0.07510 $(-1.113) $ & 0.07442 $(-1.832)$ \\
 3 & 0.07521 (1.131)*& 0.07543 $(-1.085) $ & 0.07419 $(-1.814)$ \\
 4 & 0.07502 (1.116) & 0.07497 $(-1.101) $ & 0.07476 $(-1.879)$ \\
 5 & 0.07480 (1.063) & 0.07513 $(-1.103) $ & 0.07477 $(-1.881)$ \\
 6 & 0.07486 (1.082) & 0.07490 $(-1.088) $ & 0.07476 $(-1.879)$ \\
 & & & \\
 \hline
 \end{tabular}
 \end{center}
 \vspace{2mm}

\begin{center}
\hspace{10mm} {\bf Table 2.}
 Position of the pole corresponding to the
 critical point $\kappa_c$ and residue at it
 for the indicated Pade-approximants of function
 $[\ln f_2  f_4]'$.
\vspace{2mm}

\begin{tabular}{||c|c|c|c||}
\hline
 & &  &    \\
$N$     &  $[N+1/N]$  & $[N/N]$
&  $[N/N+1]$\\
  & & & \\
  \hline
  & & & \\
 2 & 0.07418 $(-2.871)$ & 0.07461 $(-2.936) $ & 0.07558
 $(-2.963)$ \\

 3 & 0.07488 $(-2.993)$ & 0.07450 $(-2.923) $ & 0.07465
 $(-2.946)$ \\

 4 & 0.07486 $(-2.988)$ & 0.07485 $(-2.986) $ & 0.07486
 $(-2.988)$ \\

 5 & 0.07487 $(-2.989)$ & 0.07486 $(-2.987)$* & 0.07491
 $(-2.998)$* \\

 6 & 0.07481 $(-2.970)$ & 0.07484 $(-2.983) $ & 0.07483
 $(-2.978)$ \\
 & & & \\
 \hline
 \end{tabular}
 \end{center}
 \vspace{2mm}

\noindent
where the free parameter $A_0$ and trial value  for $\bar\kappa_c$
the critical point are used to accurately fit the exponent
and to  obtain a self-consistent result for $\kappa_c$ .
According to Table 3, such a fit is easy and good results for
the exponent are obtained in a wide range of the $A_0$ values.
The  optimal $A_0$  lie in the interval $0.13\div 0.63$
and a new  estimate of the critical point
$$
\kappa_c= 0.07483 \div 0.07489
$$
\begin{center}
\hspace{10mm} {\bf Table 3.}
Pade analysis of the functions
$\tilde f_i$, introduced according to Eqs.31.
  \vspace{2mm}

\begin{tabular}{||c|c|c|c||}
\hline
 & &  &    \\
$A_0$     &  $[\ln \tilde f_0(\kappa)]'$
 & $[\ln \tilde f_2(\kappa)]'$ &  $[\ln \tilde f_4(\kappa)]'$\\
     & [6/6], $\bar\kappa_c=0.074842$ & [6/6],
     $\bar\kappa_c=0.074834$ & [6/5], $\bar\kappa_c=0.074890$\\
 &   & & \\
 \hline & & &  \\
2.0 & 0.07491 (1.037)* & 0.07486
     $(-1.023) $ & 0.07488 $(-1.968)$* \\
1.0 & 0.07487 (1.018) &
     0.074844 $(-1.007) $ & 0.07493 $(-1.989)$*
     \\ 0.625 & -------- &
 0.074834 $(-1.00005) $ & --------  \\
 0.5 & 0.074855 (1.0085) & 0.07482 $(-0.996) $ & 0.07477
 $(-1.960)$ \\
 0.25 & 0.074846 (1.0029) & 0.07475 $(-0.973) $ &
 0.07488 $(-1.996) $ \\
 0.2 & -------- & --------  &   0.074890 $(-1.9994) $ \\
 0.13 & 0.0748420 (1.00005) &  --------  &  --------  \\
 0.1 & 0.074840 (0.9993) & 0.07483 $(-0.990) $* & 0.07490
 $(-2.0044)$ \\
 0.06 & 0.07487 (1.0033) & 0.07482 $(-0.988) $* & 0.07491
 $(-2.0063)$ \\
 & & & \\
 \hline
 \end{tabular}
 \end{center}
 \vspace{2mm}

\noindent
is only  slightly shifted as compared to Eq.30.  The results for
the constant $c_0$ in Eq.29 are shown in Fig.\,1a; they are close
 to the theoretical value $c_0^{th}=35.09$, but are systematically
 above  it. Similar inaccuracies in the determination of
 $c_0$ were observed in other works. The use of constants $A$,
 $B$, and $D$ for a simple hypercubic lattice from Table 5 in [7]
 gives the estimate $c_0 = B/A^2 D^4 = 142.8$ instead of
 a theoretical result of $105.2$ referring to the used
 normalization. A worse estimate was obtained in [10]; very bad
 results (discrepancies of 9 and 18 times) were obtained for
 other lattices [7]. A satisfactory test of Eq.29 was declared
 in [9], where the tested relation was not Eq.29, but its
 consequence $dg^{-1}/d\ln \tau =1/c_0$; in this case, the
 central value $c_0$ approximately corresponds to Fig.\,1a
 and the  agreement with the theory was achieved at the
 expense of an  increase in the uncertainty of the results
 because of differentiation.
\begin{figure}
\centerline{\includegraphics[width=7.2 in]{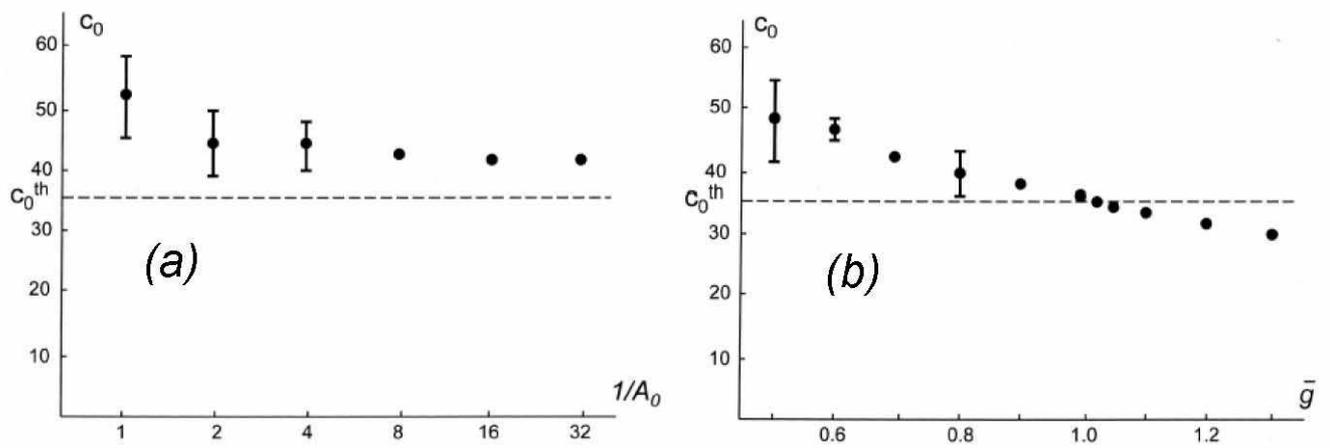}}
\caption{Constant $c_0$ in Eq.29  versus the parameter $A_0$
 in the leading logarithmic approximation (a), and  versus $\bar
 g$ in the next-to-leading logarithmic approximation (b). }
 \label{fig1}
\end{figure}

\begin{center}
{\small\bf 3.4. Second Approximation}
\end{center}

Expressions (25) and (28) are obtained in
the leading logarithmic approximation. In the
next-to-leadinglogarithmic approximation (see Appendix A),
they have the form
$$
f_0 = h_0 \,\tau \,(f_{sing})^{-p} \,,
\qquad f_2 = h_2 \,\tau^{-1}\, (f_{sing})^p \,h_{sing} \,,
\qquad f_4 = h_4 \,\tau^{-2}\, (f_{sing})^{2p-1} \,,\qquad
\eqno(33)
$$
Here, the functions $h_i(\kappa)$ are regular at $\kappa\to\kappa_c$
 and singular functions are chosen in the form
$$
f_{sing}(\kappa)=1 - \bar g \ln \tau +
        s \bar g \ln\left( 1 - \bar g \ln \tau \right)
	\,, \qquad
\eqno(34)
$$
$$
h_{sing}(\kappa)=1 + q \frac{\bar g}{f_{sing}(\kappa)} \,,
\qquad \qquad \tau = 1 - \kappa/\kappa_c
\eqno(35)
$$
where
$$
s=\,\frac{2 \beta_3}{\beta_2^2}\, - \,\frac{\zeta_1}{\beta_2}\,
=  \,\frac{n^2\!-\!8n\!-\!68}{(n\!+\!8)^2}\,,\qquad
q=\,\frac{2 \delta_2}{\beta_2^2} \,
=  \,\frac{n\!+\!2}{(n\!+\!8)^2}\,.\qquad
\eqno(36)
$$
The main distinction from Eqs.28 is reduced to the replacement
of $|\ln\tau|$ by $|\ln\tau|+s\ln|\ln\tau|$ with a known
parameter $s$; in view of the ambiguity of the normalization of
$\tau$,  it is necessary to consider the combinations
 $A+|\ln\tau|+s\ln(B+|\ln\tau|)$,
where the constants $A$ and  $B$ are different for different
functions. Formally, these constants  do not  affect
the character of a singularity, but their
unsuccessful choice can strongly distort the results. To avoid
a  large number of fitting parameters, $f_{sing}(\kappa)$
was taken in the functional form following from perturbation
theory. A reason for such a choice is as follows. The parameter
$\bar g$ has the sense of the Ginzburg number and determines the
size of the critical region, where logarithmic corrections are
significant.  It is of interest to estimate this parameter,
because the Ginzburg number is often small even in the absence
of theoretical reasons for this. The function $f_{sing}(\kappa)$
at small values of $\bar g$ is close to unity almost everywhere,
but increases sharply near $\kappa_c$. If the singularity is
separated inappropriately, regular functions $h_i(\kappa)$
in Eqs.33 are rapidly varying near $\kappa_c$  and are poorly
reproduced by Pade-approximants. However,
for small values of $\bar g$ the form of Eq.34
 is practically exact, so that functions
$h_i(\kappa)$ are almost constant. For $\bar g\agt 1$, the form
of Eq.34  is not exactly correct, but inaccuracy in the
separation of singularities in this case is not so
critical, because the function $f_{sing}(\kappa)$ is a rather
slowly varying.

The universal choice $f_{sing}(\kappa)$ for all functions is
possible if the $O(\bar g)$ contributions are
negligible as compared to unity (see Appendix A), so that the
inclusion of factors of the $h_{sing}(\kappa)$ type is strictly
speaking beyond  of accuracy. However, such factors are
sometimes of qualitative importance. In Eqs.33 they are taken
into account in the minimal  manner: the product  $f_0 f_2$
in this form has the correct singularity and ensures the correct
behavior of $\eta(g)$ at small $g$; similarly, the product
$f_4 f_2$ is incompletely free of logarithms and this property
makes it possible to slightly correct deviations
observed in Table 2.

Table 4 presents the Pade-analysis of the
functions $\tilde f_i$ introduced by the relations
$$
f_0 =\tilde f_0 \,(f_{sing})^{-p} \,,
\qquad f_2 = \tilde f_2 \, (f_{sing})^p \,h_{sing} \,,
\qquad f_4 = \tilde f_4 \, (f_{sing})^{2p-1} \,,\qquad
\eqno(37)
$$
rather than by Eqs.31; the estimate of the parameter $c_0$
in Eq.29 is illustrated in Fig.\,1b. It is easily seen that the
actual interval of $\bar g$ values  is much narrow than that in
the leading logarithmic approximation (wherethe parameter
$1/A_0$ is similar to $\bar g$). The optimum values for various
functions cover the range of $0.85\div 1.06$, which provides the
estimate
$$
c_0=36.3\pm 1.8
\eqno(38)
$$
in good agreement with a theoretical value of
$35.09$. The exact $c_0$ value is realized  at $\bar g\approx 1.02$
(see Fig.1b). Finally, Table 4 presents the maximally
accurate estimate of the critical point
$$
\kappa_c= 0.074840  \div 0.074867 \,,
\eqno(39)
$$
which is available with the existing information. The values
accepted below are $\kappa_c= 0.074850$  from the middle of interval
(39) and  $\bar g = 1.020385$, which ensures the exact $c_0$
value for the [3/3] approximant.
  \begin{center}
\hspace{10mm} {\bf Table 4.}
Pade-analysis of the functions $\tilde f_i$ introduced
 according to Eqs.37.
  \vspace{2mm}

\begin{tabular}{||c|c|c|c||}
\hline
 & &  &    \\
$\bar g$     &  $[\ln \tilde f_0(\kappa)]'$
 & $[\ln \tilde f_2(\kappa)]'$ &  $[\ln \tilde f_4(\kappa)]'$\\
     & [6/6], $\bar\kappa_c=0.074843$ & [6/6],
     $\bar\kappa_c=0.074840$ & [6/5], $\bar\kappa_c=0.074867$\\
 &   & & \\
 \hline & & &  \\
 0.5 & 0.07492 (1.036)* & 0.07488 $(-1.024) $ & 0.07487
 $(-1.968)$* \\
 0.7 & 0.07488 (1.019)* & 0.07485 $(-1.0096) $ & 0.07491
 $(-1.988)$* \\
 0.85 & -------- & 0.074840 $(-1.0008) $ & -------- \\
 0.9& 0.07485 (1.0052) & 0.074836 $(-0.998) $ & 0.074877
 $(-1.994)$ \\
  0.99& 0.074843 (1.00005) & --------  &  -------- \\
 1.0& 0.074842 (0.9995) & 0.07483 $(-0.994) $ & 0.074865
 $(-1.997)$ \\
 1.06& --------  & --------   & 0.074867
 $(-2.0001)$ \\
 1.2& 0.07482 (0.988) & 0.07476 $(-0.976) $* & 0.07488
 $(-2.010)$ \\
 & & & \\
 \hline
 \end{tabular}
 \end{center}
 \vspace{2mm}

\vspace{6mm}
\begin{center}
{\small\bf 4. RESULTS FOR RENORMALIZATION GROUP FUNCTIONS}
\end{center}

The derivatives of singular functions can be
written in the form
$$
[\ln f_{sing}]'= \frac{u_1(\tau)}{\kappa_c \tau}\,, \qquad
[\ln f_{sing}]''= \frac{u_2(\tau)}{(\kappa_c \tau)^2}\,, \qquad
[\ln h_{sing}]'= \frac{v_1(\tau)}{\kappa_c \tau}\,, \qquad
[\ln h_{sing}]''= \frac{v_2(\tau)}{(\kappa_c \tau)^2}\,,
\eqno(40)
$$
where
$$
u_1(\tau)=\,\frac{\bar g}{f_{sing}}\,
\left( 1+ \,\frac{s\bar g}{1-\bar g \ln \tau} \right) \,,
\qquad\qquad\qquad
$$
$$
u_2(\tau)=\,\frac{\bar g}{f_{sing}}\,
\left\{ 1+ \,\frac{s\bar g}{1-\bar g \ln \tau}
\,-\frac{s\bar g^2}{(1-\bar g \ln \tau)^2}\right\}
-u_1(\tau)^2 \,,
\eqno(41)
$$
$$
v_1(\tau)=\,-\frac{q\bar g^2}{f_{sing}(f_{sing}+q\bar g)}\,
\left( 1+ \,\frac{s\bar g}{1-\bar g \ln \tau} \right)\,,
\qquad\qquad\qquad
$$
$$
v_2(\tau)=\,-\frac{q\bar g^2}{f_{sing}(f_{sing}+q\bar g)}\,
\left\{ 1+ \,\frac{s\bar g}{1-\bar g \ln \tau}
\,-\frac{s\bar g^2}{(1-\bar g \ln \tau)^2}- \right.
$$
$$
\qquad\qquad\left.
-\left(\frac{\bar g}{f_{sing}} + \frac{\bar g}{f_{sing}+q\bar g}
\right)
\left( 1+ \,\frac{s\bar g}{1-\bar g \ln \tau} \right)^2
\right\}  \,.
$$
Taking into account (40), the substitution of Eqs.33
into Eqs.18 provides the parametric representation
 for the renormalization group functions in the form
 $$
g=\frac{H(\kappa)}{\kappa^2 f_{sing}}\,, \qquad
H(\kappa)=h_4 h_0^2\,,
$$
$$
\frac{\beta(g)}{g}=  \,\frac{2\kappa_c \tau
\left(2\!-\!\kappa \,[\ln h_4 h_0^2]'\right) + 2\kappa u_1  }
{ \kappa_c \tau\left(1\!-\!\kappa \,[\ln h_0]'\right)
+\kappa (1\!+\!p u_1) }   \,,
\eqno(42)
$$
$$
{    }
$$
$$
\eta(g)= \,\frac{-2\kappa_c \tau \kappa
 \,[\ln h_0 h_2]' - 2\kappa v_1  }
{ \kappa_c \tau\left(1\!-\!\kappa \,[\ln h_0]'\right)
+\kappa (1\!+\!p u_1) }   \,,
$$
$$
{    }
$$
$$
\eta_2(g)= -2\,\frac{ (\kappa_c \tau)^2
\left(1\!-\!\kappa^2 \,[\ln h_2]''\right)
+\left\{\kappa_c \tau\left(1\!+\!\kappa \,[\ln h_2]'\right)
+\kappa (1\!+\!p u_1\!+\!v_1) \right\}^2
- \kappa^2 (1\!+\!p u_2\!+\!v_2)    }
{ \left\{\kappa_c \tau\left(1\!-\!\kappa \,[\ln h_0]'\right)
+\kappa (1\!+\!p u_1) \right\}
\left\{\kappa_c \tau\left(1\!+\!\kappa \,[\ln h_2]'\right)
+\kappa (1\!+\!p u_1\!+\!v_1) \right\}
}   \,
$$
$$
{    }
$$
Asymptotic expressions (19) are obtained at $\kappa\to 0$
irrespective of the form of regular functions, whereas
at $\tau\to 0$ we have the results
$$
g=\,\frac{2\bar g}{\beta_2\,f_{sing}}\,,\qquad
\frac{\beta(g)}{g} = \,\frac{2\bar g}{f_{sing}}\,
+\,\frac{2(s-p)\bar g^2}{f_{sing}^2}\,,\qquad
\eta(g)=\,\frac{2q\bar g^2}{f_{sing}^2}\,,\qquad
\eta_2(g)=\,-\frac{2p\bar g}{f_{sing}}\,,
\eqno(43)
$$
which reproduce the first two terms of the expansion for $\beta(g)$
and the first terms of the expansions for $\eta(g)$ and
$\eta_2(g)$ in Eqs.26.\,\footnote{\,Note that the coefficients
$\beta_2$, $\beta_3$, $\delta_2$, $\zeta_1$ exhaust
invariant (scheme-independent) information on the
renormalization group functions and a further refinement of the
procedure (the construction of the next-to-next-to-leading
logarithmic approximation, etc.) requires the calculation of the
subsequent coefficients for the corresponding lattice
regularization.}
When the terms with $\tau$ are neglected, Eqs.42 provide the
regular expansions of the renormalization group functions in
$g$ (certainly without the reproduction of correct
coefficients), whereas the terms with $\tau$  provide the
$\exp(-const/g)$ singularity, which should exist owing to the
factorial divergence of the perturbation series [20, 21].
Thus, the parametric representation is  rather
"intelligent" and ensures the correct analytical properties at
$g\to 0$.

\begin{figure}
\centerline{\includegraphics[width=5.1 in]{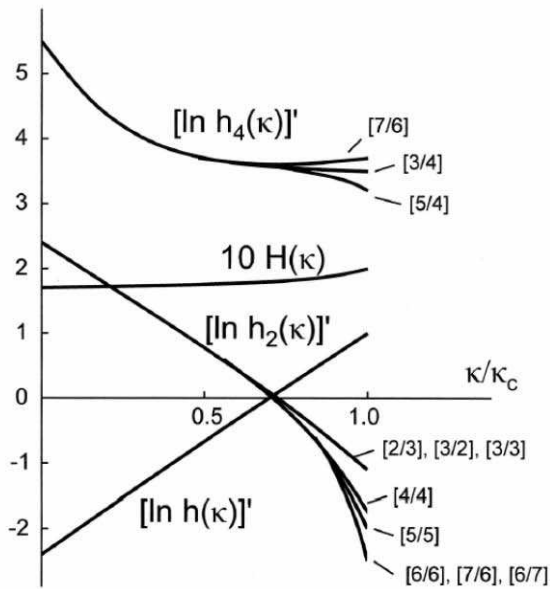}}
\caption{Regular functions $H(\kappa)$ and $[\ln h_i(\kappa)]'$
obtained in the Pade-approximation.}
\label{fig2}
\end{figure}

The accuracy of the entire construction is determined
by the accuracy of the determination of the regular functions
$h_i(\kappa)$. The expansions of these functions in $\kappa$
are obtained from Eqs.33 and are used to construct the
Pade-approximants, which are  regular in the interval
$(0,\kappa_c)$, because all singularities have been separated.
The obtained regular functions are shown in Fig.\,2. For the
functions $H(\kappa)$ and $[\ln h_0(\kappa)]'$, all
approximants  provide almost coinciding results; small
distinctions are visible for the function
$[\ln h_4(\kappa)]'$ near $\kappa_c$ (see Fig.\,2).
The situation is less satisfactory with the
function $[\ln h_2(\kappa)]'$ for which an increase in
the order of the Pade-approximation leads to an increase in the
deviations from the regular behavior predicted by lower
approximants. It is unclear whether the sequence of approximants
converged sufficiently or such deviations will further
increase. Moreover, these deviations can be artifact due to
an incompletely consistent separation of singularities
leading to a residual singularity in the function
$[\ln h_2(\kappa)]'$ (in the used approximation), which affects
higher approximants. In the
latter case, the behavior predicted by the [3/3], [2/3], and
[3/2] approximantscan be more authentic. Fortunately, this
dilemma can be resolved using the strong-coupling expansions
(see Section 5), which certainly indicate that the use of higher
Pade-approximants is correct and the results obtained in this
case are satisfactory.  Appendix B presents the parameters of the
approximants used for $H(\kappa)$  and $[\ln h_i(\kappa)]'$,
which allow the application of parametric representation (42).

To represent the results, it is convenient to use the so called
"natural normalization" of the charge, which is obtained by the
change $g\to (16\pi^2/3)g$ and corresponds to the representation
of the interaction term\,\footnote{\,The
traditional representation $g_0 \phi^4/8$ in
the $n$-component case is motivated by the fact that the vertex
$\Gamma_{\alpha\beta\gamma\delta}^{(4)}=g
I_{\alpha\beta\gamma\delta}$  in the lowest order is
$g_0 I_{\alpha\beta\gamma\delta}$, which ensures the relation
$g = g_0$ in the limit $g_0\to 0$. In
 the scalar case, the tensor $I_{\alpha\beta\gamma\delta}$  is
 reduced to three and the
 interaction is represented as $g_0\phi^4/4!$.  This motivation
 logical at first glance is in fact illusory, because the bare
 charge has no physical sense.}
in the form $(16\pi^2/4!) g_0 \phi^4$;
 in this case, the parameter $a$ in the Lipatov asymptotic form
$ ca^N\Gamma(N + b)$ [20, 21] is unity and the nearest
singularity in the Borel plane lies at the unit distance from the
coordinate origin [21]; this property defines functions varying
at an approximately unit scale. The solid lines in Fig.\,3 are the
resulting renormalization group functions, whereas the dashed
lines are the strong- and weak-coupling asymptotic behaviors.
The approach to the strong-coupling asymptotics
is strongly prolonged in agreement with the results reported in
[22]. However, the prolongation of the one-loop behavior of
the $\beta$ function pointed out in that work is not confirmed:
it appears to be an artifact, conditioned by essential
exceeding of the limiting value of $\beta(g)/g$ obtained in
[22] in comparison with Fig.\,3 [1].
\begin{figure}
\centerline{\includegraphics[width=7.2 in]{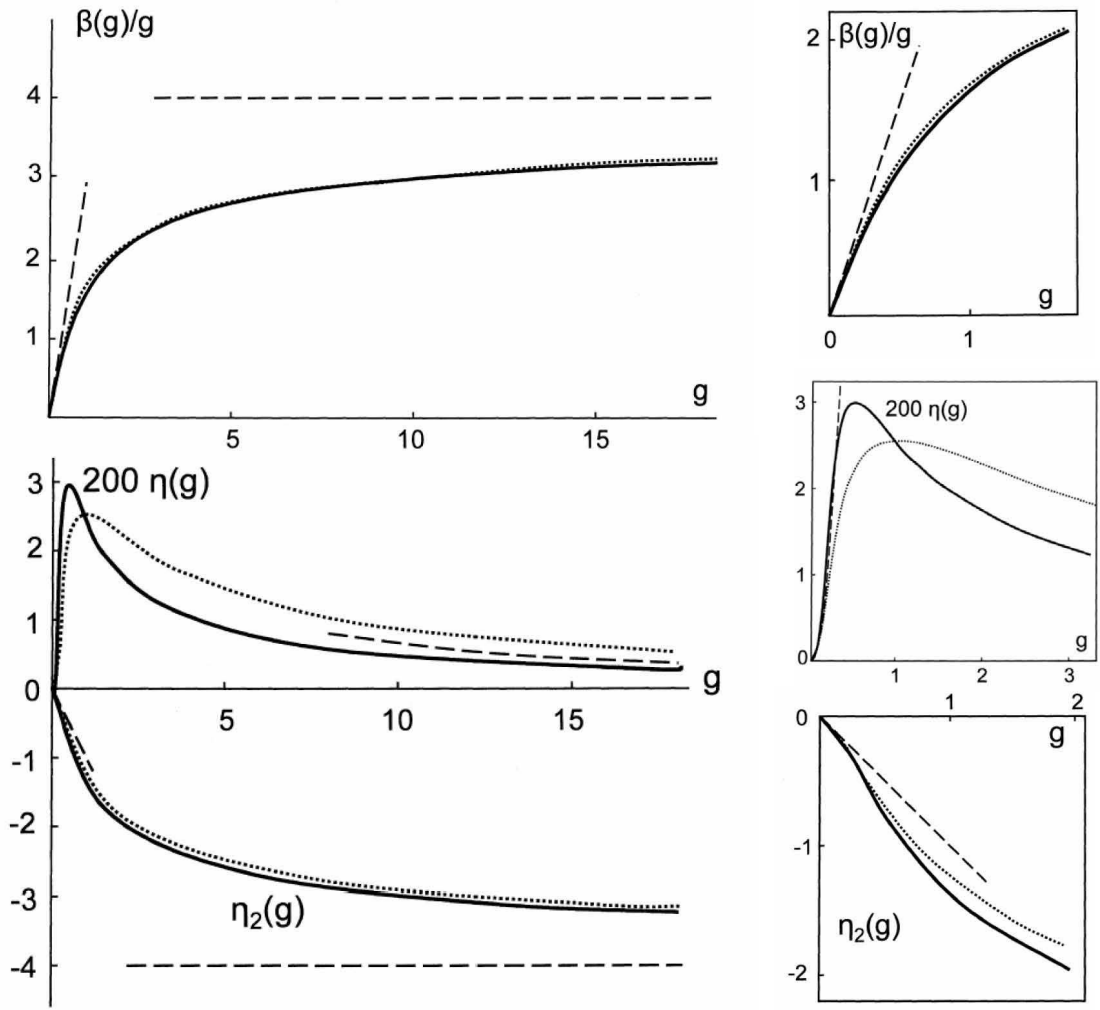}}
\caption{Solid lines are the renormalization group functions. The
dashed lines are the strong- and weak-coupling asymptotic
behaviors. The dotted lines are the results obtained under the
assumption of the constancy of regular functions $h_i(\kappa)$
under which Eqs.42 contain no information on them.}
\label{fig3}
\end{figure}

To illustrate the accuracy of the construction, the dotted lines
show  the results obtained if the functions $h_i(\kappa)$ are
changed to  constants; in this case, the
 results contain no information on these functions, because
 $[\ln h_i(\kappa)]'=0$  and a constant value of
 $H(\kappa)$  is fixed by Eq.29. It is easy to see that
 an accuracy of about 1\% for $\beta(g)/g$ and $\eta_2(g)$
 is  reached even in the complete absence of information on
 regular functions.\,\footnote{\,The reason is
 that the terms $[\ln h_i]'$ in Eqs.42 has the factor
$\kappa\kappa_c\tau=\kappa(\kappa_c-\kappa)$, which is small both for
$\kappa\to 0$ and for $\kappa\to\kappa_c$; this factor in
the middle of the interval  $\kappa=\kappa_c/2$ is
equal to $\kappa_c^2/4$, whereas the other terms are on
the order of $\kappa_c$. In
view of $\kappa_c\approx 1/15$, the effect of regular
functions on $\beta(g)/g$ and $\eta_2(g)$  is about 1\%. The
situation for $\eta(g)$ is different in view of the absentof
the $\kappa_c\tau$ term in the numerator. }

   The  real uncertainty of the construction is about
 two orders of magnitude smaller than the difference between the
 solid and dotted lines, because the regular functions (see
 Fig.\,2) are specified better than 1\% except for the region
 $\kappa>0.8 \kappa_c$, where the error for the function
 $[\ln h_2(\kappa)]'=0$  can reach 10\%.  However,
 this region corresponds to $g < 0.5$ (see Fig.\,4), where
 the effect of regular functions is insignificant.
\begin{center}
\hspace{10mm} {\bf Table 5. }
Coefficients of the expansions in $g^{-2/d}$ for
the functions $\beta(g)/g$, $\eta(g)$ and $\eta_2(g)$.
 \vspace{2mm}

\begin{tabular}{||c|c|c|c||}
\hline
 & &  &    \\
$N$     &  $\beta(g)/g$  & $\eta(g)$
&  $\eta_2(g)$\\
  & & & \\
  \hline
  & & & \\
 0 & $ $4.0000000000000 & $ $0.0000000000000 & $-$4.0000000000000 \\
 1 & $-$26.127890589687 & $ $0.0000000000000 & $ $26.127890589687 \\
 2 & $ $106.66666666666 & $ $1.7777777777777 & $-$60.444444444444 \\
 3 & $-$557.39499924665 & $-$11.612395817638 & $ $81.286770723472 \\
 4 & $ $3214.2222222221 & $ $29.708641975308 & $-$44.879012345695 \\
 5 & $-$16396.702894504 & $ $22.708685154477 & $-$1208.7213779957 \\
 6 & $ $67356.444444432 & $-$961.13125612398 & $ $9071.1992161454 \\
 7 & $-$139720.34647768 & $ $7188.4949076856 & $-$49662.878604241 \\
 8 & $-$717634.37037244 & $-$27680.892323840 & $ $197619.39191503 \\
 9 & $ $9878174.8209247 & $-$7609.7703277375 & $-$226822.08364126 \\
10 & $-$59767955.489704 & $ $938372.27840847 & $-$3873286.8465521 \\
11 & $ $186179701.36334 & $-$7226487.6363735 & $ $41826925.334797 \\
12 & $ $355069103.58896 & $ $27981910.625966 & $-$249549251.38460 \\
13 & $-$8851453360.7421 & $ $7407298.5714308 & $ $794136522.54618 \\
 & & & \\ \hline
 \end{tabular}
 \end{center}
 \vspace{2mm}

 \vspace{3mm}
\begin{center}
{\small\bf 5. STRONG-COUPLING EXPANSIONS }
\end{center}

Expanding the right-hand sides of Eqs.18  in $\kappa$ and
 expressing $\kappa$   in terms of $g$, it is easy to verify
 that the functions $\beta(g)/g$, $\eta(g)$, $\eta_2(g)$ are
 expanded in $g^{-2/d}$  as
$$
\frac{\beta(g)}{g} =\sum_{N=0}^{\infty} B_N
\left(-g^{-2/d}\right)^N\,\quad
\mbox{ \,\,\, etc.}
\eqno(44)
$$
The expansion coefficients up to $N=13$ recalculated from
high-temperature series are given in Table 5.\,\footnote{\,Fourteen
digits output by a computer are
formally presented. The accuracy decreases beginning with $N=3$
and the last four digits are unreliable at $N=13$.}

It is easy to verify that the ratios $B_{N+1}/B_N$ are the
same order of magnitude for all N, indicating the finite
convergence radius. The Pade-analysis of series (44) reveals
poles in the region $|g^{-1/2}|\sim 0.1$; these poles
for most approximants do not lie on positive semiaxis in
agreement with regularity of the renormalization group functions.
To obtain the correct power-law behavior in the limit
$g\to 0$, it is necessary to use the $[N/N+2]$ approximants for
$\beta(g)/g$ and $\eta_2(g)$  and the $[N/N+4]$ approximants for
$\eta(g)$. Such a procedure predicts $\delta_2$ with an
accuracy of about 20\%, whereas $\beta_2$ and $\zeta_1$
are estimated only by the order of magnitude. For this reason,
the summation of series (44) in the region of small $g$
gives less accurate results than the procedure described
above.

All approximants provide almost coinciding results in the
region of large $g$ ; this coincidence holds to $g = 0.5$ with
anaccuracy of about 1\%. Such
estimates for the functions $\beta(g)/g$ and $\eta_2(g)$
are in agreement with
the more accurate results obtained above. The estimates for the
function $\eta(g)$  certainly indicate that the
highest order approximants should be used for $[\ln h_2]'$ and
the results are confirmed at a level of about 1\%.  Series (44)
can apparently be used more efficiently, but analysis of
this possibility is beyond the scope of this work.

\vspace{6mm}
\begin{center}
{\small\bf 6. DISCUSSION OF THE RESULTS}
\end{center}

The resulting $\beta$ function is  non-alternating and has
the asymptotic behavior $\beta(g) = 4g$ in the
limit $g\to \infty$. According to the classification
proposed by Bogoliubov and Shirkov [23] (see discussion in [1]),
this means the possibility of the construction of a continual
theory with a finite interaction at large distances. The last
conclusion contradicts the widespread opinion that the
$\phi^4$  theory is
"trivial"  [24--28]. As was discussed in [1, 30], two
definitions --- Wilson triviality [24]  and   mathematical
triviality   [25, 26] --- were confused in the literature . The
first triviality is firmly established  (it corresponds to
positivity of the $\beta$ function), whereas pieces of
evidence in favor of the second triviality are scarce [27] and
allow another interpretation [1--30]. According to above
analysis, we have  no contradictions in the properties of
the lattice $\phi^4$ theory with the works cited in [1, 30].
However, there is a conceptual contradiction  which we want to
stress: it  concerns the role and significance of
the lattice theory.

The usual point of view  implies  that  the
lattice $\phi4$ theory provides a reasonable approximation for
the actual field theory. This interpretation provides the natural
condition $\xi\gg a$,  according to which many sites of the
lattice should be at the characteristic variation scale of the
field $\phi(x)$.  This condition can be liberalized to
$\xi\agt a$  or strengthen to $\xi/a\to\infty$. In the former case
the resriction $g\alt 1$ for renormalized charge is obtained
(for the natural normalization)  [28], while $g=0$
in the latter case (corresponding to the phase transition
point).  Thus, the usual statements are obtained:  the theory
is trivial in the continual limit ($\Lambda/m\to\infty$),
whereas in the presence of a cutoff the interaction  is limited
from above and cannot be strong. The latter circumstance is used
to obtain an upper bound for the mass of the Higgs boson [28,
29].

Our position is that the lattice theory should not
be considered   as    any approximation to  the actual
field theory
(although this is possible at $g_0\ll 1$). The continual theory
fundamentally involves no lattice; a lattice appears only in the
bare theory, which is an auxiliary construction and is
completely eliminated later. The bare theory has no physical
sense and should not satisfy any physical requirements.
Without restriction $\xi\agt a$, the renormalized charge can have
any value (see Fig.\,4). The proposed concept is completely
consistent with the "rules" accepted in mathematical works
[25, 26] according to which the continual limit $a\to 0$ is
taken at arbitrarily chosen dependences $g_0(a)$ and $m_0(a)$;
in this paper, they are taken under conditions (5).
\begin{figure}
\centerline{\includegraphics[width=5.1 in]{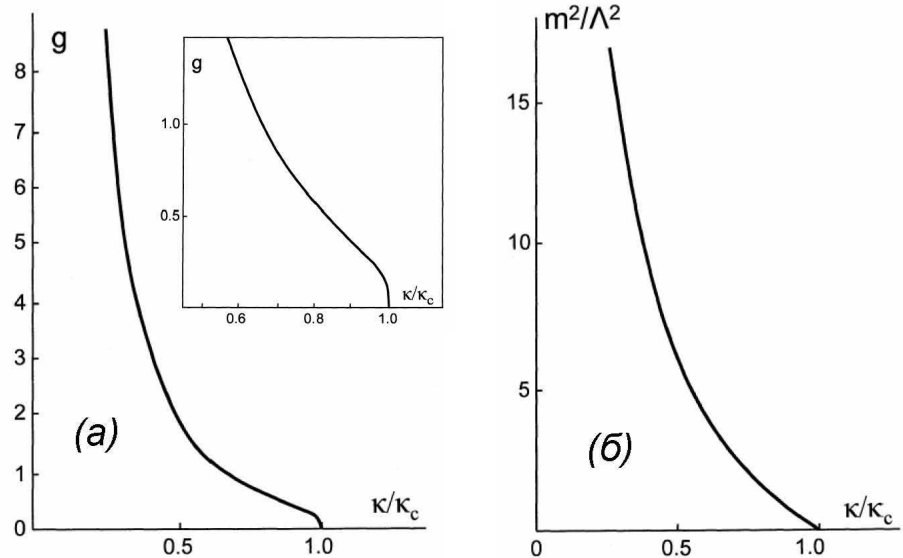}}
\caption{Renormalized  charge $g$ (a) and  mass $m$ (b) versus
$\kappa/\kappa_c$. }
\label{fig4}
\end{figure}

The only alternative for the perturbative approach is that all
quantities referring to the continual theory are expressed in
terms of functional integrals.  These integrals depend on $g_0$,
$m_0$, and $\Lambda$ and, with dimensionality taken into
account, we have for the charge, mass, and other physical
quantities $A_i$ (observables, renormalization group functions, etc.)
$$
g=F_g\left( g_0, m_0/\Lambda\right)\,,\qquad
m=\Lambda F_g\left( g_0, m_0/\Lambda\right)\,,\qquad
A_i=\Lambda^{d_i} F_i\left( g_0, m_0/\Lambda\right)\,,\qquad
\eqno(45)
$$
where $d_i$ is the  physical dimension of the quantity
$A_i$. According to Eqs.45, the real designation  of the bare
theory is to ensure the representation of the physical
quantities  in a parametric form. The relations
between $g$, $m$, and $A_i$ are of physical interest; the
parametric representation is of no deep sense in view of its
ambiguity:it can be written in various forms by changing
$g_0$  and $m_0/\Lambda$ to any other pair of variables. For this
reason, an attempt to  give  the physical sense to  the bare
theory faces the question: Why one of numerous parametrizations
is  of particular significance?

Excluding $g_0$ and $m_0/\Lambda$ in favor of
$g$ and $m/\Lambda$, it is possible to arrive at the
relation
$$
A_i=m^{d_i} \tilde F_i\left( g, m/\Lambda\right)\,.\qquad
\eqno(46)
$$
In the general case, the exclusion of the dependence on
$\Lambda$ requires the passage to the limit  $m/\Lambda\to 0$,
which corresponds to the critical point and returns us to the
"zero charge" situation. However, the central point is that
the general-position situation does not occur in Eq.46:  after
the transformation to the Ising model (valid under conditions
(5)), all functions in Eqs.45
depend on the single parameter $\kappa$; as a result,
the dependence on $m/\Lambda$ is completely
absent\,\footnote{\,This is not surprising, because the passage
to the continual limit was performed in the process of the
transformation to the Ising model [2], which was required
by the needs of renormalized, but not bare theory. }
 in Eq.(46)
$$
A_i=m^{d_i} F_i\left( g\right)\,.\qquad
\eqno(47)
$$
The renormalization program is thereby completed and no
additional passages to limits are required. This means that (a)
the lattice can be retained in the bare theory (as a convenient
technical tool for the representation of functional integrals)
and (b) the relation between $m$ and $\Lambda$ can be assumed to
be arbitrary, which ensures the attainability of any value of $g$
(see Fig.\,4).

We consider the above procedure as a real scheme for
constructing the continual $\phi^4$ theory with a finite
interaction. In fact, dependence of $g$ and $m$  on  bare
parameters (Fig.\,4), as well as the results for the
renormalization group functions (Fig.\,3), have been obtained
in the present paper.

\vspace{3mm}
\begin{center}
{\it APPENDIX A.}
{\it Next-to-Leading Logarithmic Approximation
} \end{center}

The basic  formulas referring to the next-to-leading logarithmic
approximation underlying representation (33) will be given below.
The starting point is the Callan-Symanzik equation in the cutoff
scheme\,\footnote{\,Its difference from Eq.10
is of no significance at present
context, because the first coefficients $\beta_2$, $\beta_3$,
$\delta_2$, $\zeta_1$ are independent of the renormalization
scheme. }
$$
\left[ \frac{\partial}{\partial\ln \Lambda} +
\beta(g_0) \,\frac{\partial}{\partial g_0}
 -\gamma(g_0)
\right] F\left(g_0, \Lambda/m \right) = 0\,,
\eqno(A.1)
$$
where the function $F$ satisfies the
logarithmic expansion
$$
F\left(g_0, \Lambda/m \right)
=\sum_{N=0}^{\infty}  g_0^N
\sum_{K=0}^{N}  A^K_N
\left(\ln\frac{\Lambda}{m}  \right)^K  \,.
\eqno(A.2)
$$
The substitution of (A.2) to (A.1) taking into
account the expansions
$$
\beta(g_0)=\sum_{M=2}^{\infty} \beta_M g_0^M\,,
\qquad \gamma(g_0)=\sum_{M=1}^{\infty} \gamma_M g_0^M
$$
yields the system of
recurrence relations for the coefficients $A^K_N$:
$$
-K A^K_N=\sum_{M=1}^{N-K+1}
\left[\beta_{M+1}(N-M)-\gamma_M\right] A_{N-M}^{K-1}\,,
\qquad K=1,2,\ldots,N
\eqno(A.3)
$$
In particular, for $K$ close to $N$
$$
-N A^N_N=\left[\beta_{2}(N-1)-\gamma_1\right] A_{N-1}^{N-1}\,,
\eqno(A.4)
$$
$$
-(N-1) A^{N-1}_N=
\left[\beta_{2}(N-1)-\gamma_1\right] A_{N-1}^{N-2}
+\left[\beta_{3}(N-2)-\gamma_2\right] A_{N-2}^{N-2}\,,
$$
$$
-(N-2) A^{N-2}_N=
\left[\beta_{2}(N-1)-\gamma_1\right] A_{N-1}^{N-3}
+\left[\beta_{3}(N-2)-\gamma_2\right] A_{N-2}^{N-3}
+\left[\beta_{4}(N-3)-\gamma_3\right] A_{N-3}^{N-3}\,,
$$
etc. The first equation in (A.4) is solved
immediately; after that, the next equations can be solved
one-by-one using the method of variation of constants.

{\it Vertex $\Gamma^{(1,2)}$}.
 For this vertex, $\gamma(g_0)=\eta_2(g_0)$, all coefficients
are nonzero, and  $A^0_0=1$; the first two equations in Eqs.
(A.4) give $$
A^N_N=\left(-\beta_2\right)^N\,
\frac{\Gamma(N+p)}{\Gamma(p)\Gamma(N+1)}\,,\qquad
p=-\frac{\gamma_1}{\beta_2}=-\frac{\zeta_1}{\beta_2}
\eqno(A.5)
$$
$$
A^{N-1}_N=\left(-\beta_2\right)^{N-1}\,
\frac{\Gamma(N+p)}{\Gamma(1+p)\Gamma(N)}\,
\left\{ p\,\frac{\beta_3}{\beta_2} \sum_{n=1}^{N-1}
\frac{1}{n+p} +O(1) \right\} \,.
$$
The substitution of (A.5) into (A.2) and
the summation of the corresponding series using the
formulas
$$
(1+x)^\alpha = \sum_{n=0}^{\infty}
\frac{\Gamma(n-\alpha)}{\Gamma(-\alpha)\Gamma(n+1)}\,
(-x)^n\,,\qquad
\eqno(A.6)
$$
$$
(1+x)^\alpha \ln(1+x) = \sum_{n=0}^{\infty}
\frac{\Gamma(n-\alpha)}{\Gamma(-\alpha)\Gamma(n+1)}\,
(-x)^n\, \sum_{k=0}^{n-1} \frac{1}{\alpha-k}
$$
yield
$$
\Gamma^{(1,2)}=\left\{1+O(g_0)+\beta_2 g_0\ln\frac{\Lambda}{m}
+ g_0 \frac{\beta_3}{\beta_2}
\ln\left(1+\beta_2 g_0\ln\frac{\Lambda}{m}\right)
\right\}^{-p}
\eqno(A.7)
$$
The $O(g_0)$ terms will be omitted below.

{\it The renormalized charge $g$}  satisfies Eq.(A.1) with
$\gamma(g_0)\equiv 0$, whereas all coefficients $A^N_N$ in
expansion (A.2) are zero and $A^0_1=1$.
Similar to Eqs. (A.5) and(A.7) we have a result
$$
A^{N-1}_N=\left(-\beta_2\right)^{N-1}\,,\qquad
A^{N-2}_N=\left(-\beta_2\right)^{N-2} (N-1)\,
\left\{\,\frac{\beta_3}{\beta_2} \sum_{n=1}^{N-1}
\frac{1}{n} +O(1) \right\} \,
\eqno(A.8)
$$
and
$$
g=g_0\left\{1+\beta_2 g_0\ln\frac{\Lambda}{m}
+ g_0 \frac{\beta_3}{\beta_2}
\ln\left(1+\beta_2 g_0\ln\frac{\Lambda}{m}\right)
\right\}^{-1}
\eqno(A.9)
$$
which can also be obtained directly from the GellMann - Low
equation.

{\it Renormalized mass.}. Neglecting the $Z$ factor, the
Ward identity
$$
\Gamma^{(1,2)}= \frac{d}{d m_0^2}\,\Gamma^{(0,2)}=
\frac{d}{d m_0^2}\,\frac{m^2}{Z}
\eqno(A.10)
$$
can be written in the form $dm_0^2/dm^2=1/\Gamma^{(1,2)} $;
the integration with respect to $m^2$ within the necessary
accuracy is reduced to the multiplication by $m^2$,
$$
m^2=(m_0^2-m_c^2) \left\{1+\beta_2 g_0\ln\frac{\Lambda}{m}
+ g_0 \frac{\beta_3}{\beta_2}
\ln\left(1+\beta_2 g_0\ln\frac{\Lambda}{m}\right)
\right\}^{-p} \,,
\eqno(A.11)
$$
where $m_c^2$  is the value of  $m_0^2$ corresponding to the
transition point. The introduction of the dimensionless distance
to the transition $\tau\propto (m_0^2-m_c^2)$  and iterative
exclusion of $m$ from the righ-thand side give
$$
m^2=\tau \left[1+\bar g\ln 1/\tau
+ s\bar g \ln\left(1+\bar g\ln 1/\tau \right)
\right]^{-p} \,, \qquad \bar g=\beta_2 g_0/2
\eqno(A.12)
$$
where $s$ is given in Eq.36. Similarly, (A.9) reduces to
the form
$$
g=\frac{2}{\beta_2} \frac{\bar g}{1+\bar g\ln
1/\tau + s\bar g \ln\left(1+\bar g\ln 1/\tau \right) }
\,.
\eqno(A.13)
$$

\vspace{2mm}

{\it The $Z$ factor} satisfies Eq. (A.1) with
$\gamma(g_0)=-\eta(g_0)$, while $A^0_0=1$, $A^0_1=A^1_1=0$ and
$A^N_N=0$ for $N\ge 2$ in expansion (A.2).  Similar to  (A.8),
we have for $N\ge 2$
$$
A^{N-1}_N=A^1_2\left(-\beta_2\right)^{N-2}\,,\qquad
A^{N-2}_N=A^1_2 \left(-\beta_2\right)^{N-2} (N-1)\,
\left\{\,-\frac{\beta_3}{\beta_2^2} \sum_{n=2}^{N-1}
\frac{1}{n} +O(1) \right\} \,,
\eqno(A.14)
$$
and after summation
$$
Z=1+\frac{A^1_2 g_0}{\beta_2} -\frac{A^1_2 g_0}{\beta_2}
\left\{1+\beta_2 g_0\ln\frac{\Lambda}{m}
+ g_0 \frac{\beta_3}{\beta_2}
\ln\left(1+\beta_2 g_0\ln\frac{\Lambda}{m}\right)
\right\}^{-1}
\eqno(A.15)
$$
Taking into account the relation
$A^1_2=-\delta_2$, expressing $m$ in terms of $\tau$
and omitting an insignificant constant factor,
one obtains with the
necessary accuracy
$$
Z=1+\frac{2\delta_2 }{\beta_2^2} \frac{\bar g}{1+\bar g\ln
1/\tau + s\bar g \ln\left(1+\bar g\ln 1/\tau \right) }
\,.
\eqno(A.16)
$$
The substitution of (A.12), (A.13), (A.16) into the relations
$$
\frac{K_2}{\tilde K_2}=m^2\,,\qquad
\frac{K_2}{ K_0}= \frac{Z}{m^2} \,,\qquad
\frac{K_4 K_0}{ K_2^2}= -\frac{g}{m^4} \,,
\eqno(A.17)
$$
yields Eqs.33 for $f_i(\kappa)$. The difference of the
$Z$ factor from unity corresponds to the corrections of the
order $g_0/ln\tau$, which were neglected above, and
strictly speaking is beyond the accuracy.  However, without
the inclusion of the $Z$ factor, the product $f_0 f_2$ would be
a regular function and, correspondingly, the behavior of
$\eta(g)$ at small $g$ values would be incorrect. For this
reason, the function $h_{sing}$ corresponding to the Z factor
isintroduced in Eqs.33 by the minimal  manner to ensure the
correct singularity in $f_0 f_2$.

\begin{center}
\hspace{0mm} {\bf Table 6.}
Parameters of Pade-approximation (24) of regular functions \\ \vspace{3mm}

\hspace{17mm}$H(\kappa)$  \hspace{30mm}   $[\ln h_0(\kappa)]'$
\begin{tabular}{|c||c|c||c|c||}
\hline

$n$ & $p_n$ &$q_n$   & $p_n$ &  $q_n$  \\
    \hline
&   &   &  &    \\
0&  0.166666  &    1.000000  &    -2.389114     & 1.000000  \\
1&  2.173343  &    12.28756  &     39.93594     & 1.218909  \\
2&$-8.874246$ &  $-6.056224$ &     134.2565     &$-14.76806$ \\
3&  103.5876  &  $-124.8396$ &   $-1759.943$    & 498.1762  \\
4&  0         &       0      &     14434.97     &$-2468.179$ \\
 &  &   &  & \\
\hline \end{tabular} \end{center}

\begin{center}
\hspace{17mm}$[\ln h_2(\kappa)]'$  \hspace{25mm}   $[\ln
h_4(\kappa)]'$ \begin{tabular}{|c||c|c||c|c||}
\hline
$n$ & $p_n$ &$q_n$   & $p_n$ &  $q_n$  \\
    \hline
&   &   &  &  \\
0&   2.416517   &   1.000000  &   5.530725 &    1.000000 \\
1& $-50.63241$  & $-3.794992$ &   13.37787 &    21.09480 \\
2& $-345.9676$  & $-201.7335$ &   630.6971 &    57.28333 \\
3&   9156.772   &   738.3887  &   3430.220 &    252.1934 \\
4& $-1285.833$  &  4787.275   &   0        &    10511.06 \\
5& $-267488.9$  & $-26827.13$ &   0        &    0        \\
6&   109199.7   &   363530.4  &   0        &    0         \\
 &  &   &  & \\
\hline \end{tabular} \end{center}

\vspace{3mm}
\begin{center}
{\it APPENDIX B.} {\it Pade-Approximation of Regular Functions
} \end{center}

 Table 6 shows the  coefficients $p_n$ and $q_n$ in Eq.24
 for  the Pade-approximantion of the regular functions
 $H(\kappa)$ and $ [\ln h_i(\kappa)]'$; the lowest order
 approximants having the complete accuracy are presented.
 The  singularities were separated with the values
 $\kappa_c = 0.074850$ and $\bar g =  1.020385$.

\end{document}